\documentclass{emulateapj}

\shorttitle{Curvature effects}
\shortauthors{M. Luna et al.}

\begin{document}

\title{The effects of magnetic-field geometry on longitudinal oscillations of solar prominences}

\author{M. Luna\altaffilmark{1}, A. J. D\'\i az\altaffilmark{3,4}, and J.
Karpen\altaffilmark{2}}

\altaffiltext{1}{CRESST and Space Weather Laboratory NASA/GSFC, Greenbelt, MD 20771, USA}
\altaffiltext{2}{NASA/GSFC, Greenbelt, MD 20771, USA}
\altaffiltext{3}{Instituto de Astrof\'{\i}sica de Canarias, E-38200 La Laguna, Tenerife, Spain}
\altaffiltext{4}{Departamento de Astrof\'{\i}sica, Universidad de La Laguna, E-38206 La Laguna, Tenerife, Spain}

\begin{abstract}

We investigate the influence of the geometry of the solar filament magnetic structure on the large-amplitude longitudinal oscillations. A representative filament flux tube is modeled as composed of a cool thread centered in a dipped part with hot coronal regions on either side. We have found the normal modes of the system, and establish that the observed longitudinal oscillations are well described with the fundamental mode. For small and intermediate curvature radii and moderate to large density contrast between the prominence and the corona, the main restoring force is the solar gravity. In this full wave description of the oscillation a simple expression for the oscillation frequencies is derived in which the pressure-driven term introduces a small correction. We have also found that the normal modes are almost independent of the geometry of the hot regions of the tube. We conclude that observed large-amplitude longitudinal oscillations are driven by the projected gravity along the flux tubes, and are strongly influenced by the curvature of the dips of the magnetic field in which the threads reside.
\end{abstract}

\section{Introduction}\label{intro-sec}

Large-amplitude longitudinal (LAL) oscillations in prominences were first reported by \citet{jing2003}; since then, only few additional reports of these motions have appeared \citep{jing2006,vrsnak2007,zhang2012}. These oscillations produce motions along the magnetic field, have long periods of 50-160 minutes, and are damped in $2.3$-$6.2$ cycles, with high velocity amplitudes in the range $30$-$100~\mathrm{km\,s^{-1}}$. LAL oscillations are apparently triggered by an energetic event: a sub-flare, a microflare, or a flare close to the filament.

Several models have been proposed to explain the restoring force and damping mechanism of the LAL oscillations \citep[see review by][]{tripathi2009}, but most do not successfully describe the thread motions. Recently, we studied the oscillations of threads forming the basic components of a prominence \citep{luna2012a} in a 3D sheared arcade \citep{devore2005,luna2012}. We found that the restoring force is mainly the gravity and the pressure forces are small. This type of oscillation resembles the motion of a gravity-driven pendulum, where the frequency only depends on the solar gravity and the flux-tube dip curvature. We estimated the minimum value of the magnetic field at the tube dips and found agreement with previous estimates and observed values. Additionally, this study revealed a new method for measuring the radius of curvature of the filament dips. \citet{zhang2012} observed and analyzed an oscillating prominence and found that the motion is produced along a dipped magnetic field, in agreement with \citet{luna2012a}. These studies reveal that the LAL oscillations are strongly related to the filament-channel geometry.

The filament-channel structure is not well understood and several models have been suggested. The sheared arcade and the flux rope models are the most successful candidates explaining most of the observational evidence \citep[see the review by][]{mackay2010}. In these models the magnetic structure is static and independent of the prominence evolution because the plasma-$\beta$ is small, and the structure has dips where the cool prominence resides. Due to the low plasma $\beta$, however, most models agree that these dips are not caused by weight of the prominence. The prominence mass forms in the dipped part of the magnetic field because it is a gravity potential well where evaporated mass plasma condense and collect \citep{antiochos1999,karpen2003}. There is also direct observational evidence from polarimetric inversions of the dipped magnetic structure of the filaments \citep{lass06,slsl12}. In these models the curvature in the dips is large and the LAL oscillations could be strongly influenced by these geometries. In contrast, where the magnetic field is slightly curved by the prominence weight, the curvature effects on the oscillations are negligible \citep{oliver1992,oliver1993,oliver1995,terradas2001}.

In our previous study of LAL oscillations, we assumed the threads to be solid masses moving in curved flux-tubes without interaction with the surrounding hot plasma. In this work we use a full wave description of the oscillation, expanding our previous investigations. We focus on the restoring forces of the LAL oscillations and the influence of the curvature of the filament magnetic fields in different tube geometries, and compare the resulting thread motions with observed LAL oscillations properties.

\section{Flux tubes with curvature}\label{tubes-curv-sec}

In this work we assume that the plasma is low-$\beta$ and confined with static magnetic field. In this regime the plasma motion is described with Equations (1)-(4) of \citet{karpen2005}. We additionally consider that the system is adiabatic, with no heating and radiation, and the tubes have a uniform width. Thus, the terms associated with the energy loss and gains, and the area expansion, can be neglected. We linearize this set of equations to obtain the equations for the perturbed quantities
\begin{eqnarray}\label{linear-eq1}
\frac{\partial \rho_{1}}{\partial t} + v \frac{\partial \rho_{0}}{\partial s} + \rho_{0} \frac{\partial v}{\partial s} &=& 0 ~,\\ \label{linear-eq2}
\rho_{0} \frac{\partial v}{\partial t} + \frac{\partial p_{1}}{\partial s} - \rho_{1} g_{\parallel} &=& 0 ~,\\ \label{linear-eq3}
\frac{\partial p_{1}}{\partial t} + v \frac{\partial p_{0}}{\partial s} + \gamma p_{0} \frac{\partial v}{\partial s} &=& 0~,
\end{eqnarray}
where the ``$0$'' index means the equilibrium quantity that depends on the coordinate s, the ``$1$'' index means the perturbed quantity, $v$ is the perturbed velocity (a zero background velocity is considered), and $g_{\parallel}$ is the gravity projected along the flux tube. Additionally the plasma is in hydrostatic equilibrium
\begin{equation}\label{mechequil-eq}
\frac{\partial p_{0}}{\partial s} = \rho_{0} g_{\parallel}~.
\end{equation}
Combining Equations (\ref{linear-eq1})-(\ref{mechequil-eq}) we obtain the equation for the velocity perturbation
\begin{equation}\label{wave-eq}
\frac{\partial^{2} v}{\partial t^{2}} - c_{s}^{2} \frac{\partial^{2} v}{\partial s^{2}} =\gamma g_{\parallel} \frac{\partial v}{\partial s} + v \frac{\partial g_{\parallel}}{\partial s}~,
\end{equation}\\
where $c_{s}=c_{s}(s)=\gamma p_{0}(s)/\rho_{0}(s)$. This equation is similar to Equation (7.30) of \citet{goedbloed2004} assuming a displacement constrained along the flux tube.

We model the filament flux-tube geometry as composed of up to 3 curved segments
(see Fig. \ref{sketch}). These segments are contained in a vertical plane with the center of curvature located above or below the tube for a concave-up or concave-down segments respectively. Each segment has a constant radius of curvature $R$. The radius of curvature is positive for a concave-up segment and negative
for a concave-down segment. With these considerations, on
each segment of the tube Equation (\ref{wave-eq}) is 
\begin{equation}
\frac{\partial^{2} v}{\partial t^{2}} - c_{s}^{2} \frac{\partial^{2} v}{\partial s^{2}} =-\gamma g_{0} \sin \theta \frac{\partial v}{\partial s} - v \frac{g_{0} \cos \theta}{R}~,
\end{equation}
where $1/R=\partial \theta / \partial s$ defining $\theta = s/R$. Assuming now that the radius of curvature
$R$ is sufficiently large to fulfill the condition $|s/R| \ll 1$, the $\theta$ angle takes small values in all the
positions of the tube. Thus considering linear approximations of the sinusoidal
functions ($\sin \theta = \theta$ and $\cos \theta = 1$), then
\begin{equation}\label{motion-eq}
\frac{\partial^{2} v}{\partial t^{2}} - c_{s}^{2} \frac{\partial^{2} v}{\partial
s^{2}} +\frac{\gamma g_{0} s}{R} \frac{\partial v}{\partial s} + v
\frac{g_{0}}{R} = 0~.
\label{pde_curv}
\end{equation}
This equation reduces to Equation~(8) of \citet{diaz2006} for a low-$\beta$ plasma when curvature
is negligible (i.e., $R \to \infty$). Hence, for a straight tube the
equation describing the perturbations is simply
\begin{equation}
\frac{\partial^{2} v}{\partial t^{2}} - c_{s}^{2} \frac{\partial^{2} v}{\partial
s^{2}} =0.
\label{straightwave_eq}
\end{equation}

We consider solutions for which the time dependence is a simple harmonic oscillation with frequency $\omega$ and the velocity perturbation takes the form $v(s) e^{-i \omega t}$. Thus Equation (\ref{motion-eq}) becomes
\begin{equation}\label{wavegravity-eq}
c_{s}^{2} \frac{\partial^{2} v}{\partial s^{2}} - \frac{\gamma g_{0} s}{R} \frac{\partial v}{\partial s} + \left( \omega^{2}-  \frac{g_{0}}{R} \right) v = 0~.
\end{equation}
This equation describes the oscillatory motion of a plasma embedded in a curved tube with constant curvature in hydrostatic equilibrium. 

\subsection{Isothermal plasma flux-tube segments}

Solving Equation (\ref{wavegravity-eq}) is complicated because the sound velocity depends on the position along the tube, i.e. $c_{s}=c_{s}(s)$. However, if we assume that the segments of the tube are isothermal, the sound velocity is uniform on each segment, i.e. $c_{s} \neq c_{s}(s)$. Thus, we can perform a linear change of variable defined by
\begin{equation}
r=s \left( \frac{\gamma g_{0}}{2 c_\mathrm{s}^{2} R} \right)^{1/2}~,
\label{chn_var}
\end{equation}
and Equation~\ref{wavegravity-eq} becomes
\begin{equation}\label{waveadim-eq}
 \frac{\partial^{2} v}{\partial r^{2}} -2 r \frac{\partial v}{\partial r}  +
 \frac{2}{\gamma}\left(\frac{R \omega^2}{g_0} -1 \right) v =0.
\end{equation}

We consider that the plasma within the flux tubes of the filament is distributed in three isothermal regions. The thread consists of cool plasma centered at the dip with an uniform sound speed, $c_\mathrm{s p}$. The remaining thermal regions are the hot coronal plasma filling the tube from both ends of the thread to the footpoints, with a uniform sound speed, $c_\mathrm{s c}$. In this work we ignore the small transition region from the cool prominence thread to the coronal hot plasma (the so-called PCTR) at both ends of the thread. This region is thin in comparison with the thread length, and the influence on the thread oscillations may be small, so we leave the incorporation of this region for a future study. We note that both the equilibrium density and pressure depend on $s$, despite having a uniform sound speed along each flux-tube segment. In this situation the density and pressure decrease exponentially with height, at a rate given by the pressure scale height. For a typical hot coronal plasma the pressure scale height is a few tens of $\mathrm{Mm}$. However, for a typical prominence it is about $0.2~\mathrm{Mm}$. Therefore the pressure and density are roughly uniform in the coronal parts of our low-lying flux tubes, whereas they vary rapidly along the curved thread.

Equation (\ref{waveadim-eq}) governs the plasma motion on each part of the piecewise flux tube with uniform temperature and $R$. This equation is a second-order
ordinary equation with a irregular singular point at infinity, so its solution
can be expressed as a combination of two regular linearly independent functions
at each point (except at infinity). The general solution can be written in terms
of confluent hypergeometric functions (CHFs) $M(a,b;x)$ 
\citep[see][]{abramowitz1972} as
\begin{eqnarray}
v (r) &=& A_1 \, M \left( -\frac{\lambda}{4}, \frac{1}{2}; r^2 \right)  \nonumber \\
&+& A_2 \, r \, M \left( -\frac{\lambda-2}{4}, \frac{3}{2}; r^2 \right),
\label{gen_sol}
\end{eqnarray}

\noindent
with $A_1$ and $A_2$ being two arbitrary constants and
\begin{equation}
\lambda= \frac{2}{\gamma} \left(\frac{R \omega^2}{g_0}-1 \right) \, .
\label{lambda_def}
\end{equation}
This solution can be expressed in terms of the well-known Hermite polynomials if
$\lambda=2 n$, with $n$ being an integer, but even in this case there is a
second regular solution which cannot be discarded. Hence, it is suitable to work
directly with the general solution in terms of CHFs (Eq. \ref{gen_sol}). 

Regarding the symmetry at $r=0$, the CHFs have the following property at the origin:
\begin{equation}
M \left( a,b,0 \right) = 1,
\end{equation}
\noindent
so we obtain a symmetric function by setting $A_2=0$ and an antisymmetric one
by setting $A_1=0$.

\begin{figure}[!h]
  \center{\resizebox{\hsize}{!}{\includegraphics{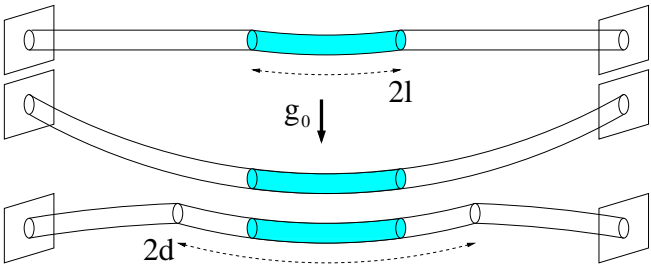} }}
\caption{Sketch of the configurations considered in this study, differentiated by
the shape of the flux tube in the hot plasma region. The upper sketch
corresponds to Model 1, the middle sketch to Model 2, and the lower sketch to the
M-shaped Model 3. The length of the flux tubes is $2L$, the length of thread is $2 l$, and the dipped part of the tube has a length $2 d$ in Model 3. }
\label{sketch}
\end{figure}

In this work we use three flux-tube models to investigate the influence of the hot region on the motion of the
cool plasma. The cool thread is clearly located in a concave-up field depression or dip as shown in our earlier study \citep{luna2012}. Model 1 is the simplest: a tube with straight hot regions connecting to the chromosphere (see upper sketch of Fig. \ref{sketch}). Model 2 is a tube which is concave-up in all domains (see middle sketch of Fig.~\ref{sketch}b). Finally, Model 3 has the most complex geometry considered in this work, as shown in the bottom sketch of Figure~\ref{sketch}: a dipped part containing a central cool thread and two hot plasma regions at both sides, connected with two concave-down segments forming the legs of a M-shaped tube.

\subsection{Boundary conditions}\label{boundaryconditions-sec}

The solutions of each region of our piecewise flux tube model must be joined by the boundary conditions. The resulting solutions are the normal modes. To solve this problem we need two types of boundary conditions:  line-tying at the chromospheric footpoints of the tube (located at $s=\pm L$, with $2L$ being the length of the supporting magnetic flux tube) and jump conditions at the plasma interfaces. We adopt the simplest possible conditions at the footpoints: a line-tied rigid wall with no flow \citep{hood1986,van-der-linden1994,diaz2004}, namely
\begin{equation}
v(\pm L)=0~.
\label{line-tying}
\end{equation}
Since the chromosphere is much denser than the corona, it behaves as a purely reflecting layer. This condition is appropriate mainly because there is not enough energy in the perturbations carried by the coronal material to drive important motions in the chromosphere. Moreover, this simple condition completely decouples the coronal and the photospheric plasmas (so the dynamics of the photospheric plasma are not taken into account), although we plan to address more advanced ``flow-through" conditions and coupling with the chromosphere in future works.

The second set of boundary conditions is prescribed at the interface between the cool and hot plasmas, located at $s=l$, with $2l$ being the length of the dense thread. We need to carefully deduce these boundary conditions. Following \citet{chandrasekhar} we integrate the differential Equations (\ref{linear-eq1})-(\ref{linear-eq3}) across the boundary (from $s=l-a$ to $s=l+a$) and let $a\to 0$. The integral of any variable that does not have an infinite jump becomes zero in that limit. First, integrating the hydrostatic equilibrium (Eq. \ref{mechequil-eq}) we obtain $[p_0]=0$ with the commonly used notation $[a]=a_2-a_1$ for the jump between mediums $1$ and $2$. Therefore the equilibrium pressure is continuous at the boundary. Similarly, from Equations~(\ref{linear-eq1})-(\ref{linear-eq3}) the two boundary conditions for the velocity are
\begin{equation}
[v]_{s=l}=0, \,\,\,\,  \left[\frac{dv}{ds} 
\right]_{s=l}=0~.
\label{jump_cond}
\end{equation}
Thus, including gravity does not affect the jump conditions, in agreement with \citet{diaz2004} and \citet{diaz2006}.

\section{Uncoupled thread}\label{uncoupled-sec}

In a first approximation we consider the oscillation of the thread alone, assuming that the hot plasma filling the rest of the tube is irrelevant to the dynamics of the prominence. This assumption is justified by our previous work \citep{luna2012a}, in which we found that the dynamics of the threads are basically governed by the gravity, and the interaction with the ambient hot plasma is small. In this situation the boundary conditions of Equation (\ref{jump_cond}) are not applicable. We assume that the thread moves freely and thus the boundaries at $s=\pm l$ are open. The motion of the thread is described by Equation (\ref{gen_sol}); the $A_{1}$, $A_{2}$, and $\lambda$ parameters can be chosen freely because there is no constraining condition. 

\begin{figure}[h]
  \center{\resizebox{\hsize}{!}{\includegraphics{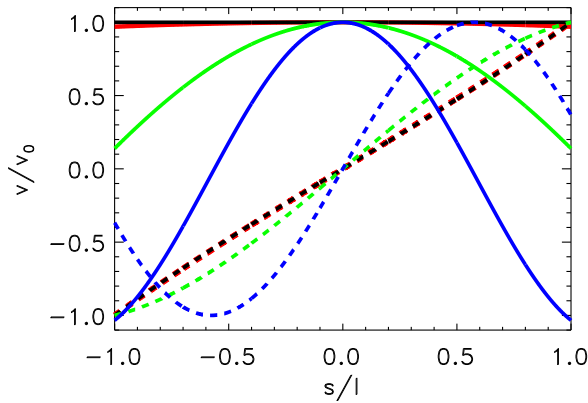} }}
\caption{The symmetric (solid lines) and antisymmetric (dashed lines) 
normalized solutions of Eq. (\ref{gen_sol}) for the values $\lambda=0$ (black), $\lambda=0.3$ (red), $\lambda=10$ (green), and $\lambda=40$ (blue) for a thread of half-length $l= 5~\mathrm{Mm}$, 
$R=75~\mathrm{Mm}$, and $c_\mathrm{sp}=20~\mathrm{km\,s^{-1}}$.}
\label{chf-plots-fig}
\end{figure}

In Figure \ref{chf-plots-fig}, we have plotted different symmetric and antisymmetric solutions of Equation (\ref{gen_sol}) along the thread, $|s|\le l$, for different $\lambda$ values.
For a symmetric solution and $\lambda=0$ the whole thread moves as a rigid body with an uniform velocity. For a small value ($\lambda=0.3$) the symmetric velocity distribution differs only slightly from the uniform case: the velocity at the ends of the thread is slightly smaller than in the center. For $\lambda=10$ the velocity clearly is not uniform but still positive. For $\lambda=40$ the motion is more complex, however to with the thread ends moving out of phase with respect the center of the thread. Hence, the $\lambda$ parameter gives the spatial coherence of the velocity in the thread. For small $\lambda$ the spatial coherence is high, with a uniform velocity distribution, whereas for larger values the spatial coherence is small with steep velocity gradients along the thread. For antisymmetric solutions and $\lambda=0$ both ends of the threads move in opposite directions, compressing and rarefying the thread plasma, whereas the center remains at rest. For $\lambda=0.3$ the antisymmetric solution is slightly different, whereas the solution is more complicated for $\lambda=10$ and $40$.

The general solution is a superposition of these modes, where the exact combination depends on the initial conditions. One special case is worth studying carefully: an exact rigid-body perturbation in which the whole thread is shifted from the initial equilibrium position. In this case the solution is symmetric and the spatial coherence parameter is $\lambda=0$. Thus the frequency of the oscillation is 
\begin{equation}
\omega_\mathrm{g}=\left( \frac{g_0}{R} \right)^{1/2}~.
\label{pendulum}
\end{equation}
This solution was described in \citet[][Eq. 4]{luna2012a}, and predicts the thread oscillating as a gravity-driven pendulum. For a general initial perturbation $A_0(s,t=0)$ this solution is no longer valid and Equation (\ref{pde_curv}) must be solved instead.

\section{Curved thread with a straight field in the hot plasma region}\label{model1-sec}

We consider now Model 1, where the thread is located in the center of the flux-tube dip
and the remainder of the flux tube is straight and filled with hot coronal plasma (upper sketch of Fig.~\ref{sketch}). The solution to this piecewise model is given by Equation~(\ref{gen_sol}) in the dense region and Equation~(\ref{straightwave_eq}) with $g_0/R \to 0$ in the evacuated region, namely
\begin{equation}
v(s)=\left\{ \begin{array}{ll}
    B_{1} \, M \left( -\frac{\lambda}{4}, \frac{1}{2}; r_\mathrm{p}^2 \right) + \\
    +B_{2} \, r_p \, M \left( -\frac{\lambda-2}{4}, \frac{3}{2}; r_\mathrm{p}^2 \right), 
    			& 0 < s < l , \\
    D_{1} \, \sin  \left\{ \frac{\omega}{c_\mathrm{sc}} (L-s) \right\}, 
			& l < s < L . \end{array} \right.
\label{sin_sol}
\end{equation}
We have ensured that the line-tying boundary condition (Eq. \ref{line-tying}) is enforced by choosing only the sine function in the hot region. Using the series expansion for the CHF, it can be proved that
\begin{equation}
\lim_{g_0/R \to 0} M\left( \frac{g_0/R-\omega^2}{2 \gamma g_0/R}, \frac{1}{2}; 
	s^2 \frac{\gamma g_0}{2 c_\mathrm{sp}^2 R} \right) = 
	\cos{\frac{\omega s}{ c_\mathrm{sp}}},
\end{equation}
so we recover the fully-straightened case with very large $R$ studied in \citet{diaz2002,diaz2010}.

Next, we apply the boundary conditions (Eq. \ref{jump_cond}) at the interface
between the cool and hot plasma at $s=\pm l$. We
obtain the following dispersion relation for the symmetric normal modes  (symmetric with respect to $s=0$) after eliminating the amplitude constants $B_1$ and $D_1$, and considering $B_2=0$:
\begin{eqnarray}
-\frac{\omega}{c_\mathrm{sc}} \mathrm{cotg}&&\!\!\!\!\!\!\left[ \frac{\omega}
	{c_\mathrm{sc}} (L-l) \right] = - \frac{l g_0}{c_\mathrm{sp}^2 R} 
	 \left(\frac{\omega^2 R}{g_0} -1\right)  \nonumber \\
	&\times&\frac{M\left( 1+\frac{1}{2\gamma}-\frac{\omega^2 R}{2 \gamma g_0},
	\frac{3}{2}; 
	\frac{\gamma g_0 l^2}{2 R c_\mathrm{sp}^2} \right)}{M\left(\frac{1}{2\gamma} - 
	\frac{\omega^2 R}{2 \gamma g_0},\frac{1}{2}; 
	\frac{\gamma g_0 l^2}{2 R c_\mathrm{sp}^2}\right)}~.
\label{dr_1}
\end{eqnarray}

We define a new set of dimensionless variables as
\begin{equation}
\Omega=\frac{\omega}{\omega_{\mathrm{g}}},~W=\frac{l}{L}, ~\Phi=\frac{\omega_{\mathrm{g}} L}{c_{s c}}, ~\chi=\frac{c^{2}_\mathrm{sc}}{c^{2}_\mathrm{sp}}~.
\label{dim_units}
\end{equation}
With these definitions the dispersion relation takes the form
\begin{eqnarray}\nonumber 
\Omega \, \mathrm{cotg}\left[\Omega \Phi (1-W) \right]= \chi W \Phi \left( \Omega^2-1 \right)\\
\times\frac{ M\left( 1+\frac{1}{2\gamma} -\frac{\Omega^2}{2\gamma},
\frac{3}{2}; \frac{\chi \gamma W^2 \Phi^2}{2}\right)}{
 M\left( \frac{1}{2\gamma} -\frac{\Omega^2}{2\gamma},\frac{1}{2}; 
\frac{\chi \gamma W^2 \Phi^2}{2}\right)}~.
\label{drm1sym-eq}
\end{eqnarray}
The parameter $\chi$ is usually called density contrast when the density and pressure are assumed uniform in each segment along the tube. However, in this work we consider nonuniform densities and pressures. The variable $\chi$ is equal to the density contrast only at the interface $s=l$ because the pressure must be continuous across the thread-corona interface (see \S \ref{boundaryconditions-sec}). Thus $\chi=\rho_\mathrm{p}(l)/\rho_\mathrm{c}(l)$ and we call this parameter the contrast or density contrast hereafter.
\begin{figure}[h!]
  \center{\resizebox{\hsize}{!}{\includegraphics{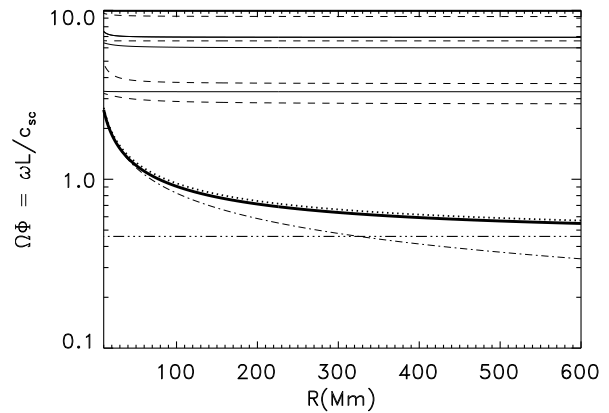} }}
\caption{The symmetric (solid lines) and antisymmetric (dashed lines) normal modes for Model 1 as a function of the radius of curvature $R$ of the dip. The fundamental mode is plotted with a thick solid line. The gravity-driven frequency, $\omega_\mathrm{g}$, defined by Eq. \ref{pendulum} (dot-dashed line), and the pressure-driven frequency, $\omega_\mathrm{s}$, of Eq. \ref{w-pressuredriven-eq} (three dot dashed line), are also plotted. Additionally, the approximate fundamental frequency of Eq. \ref{wtotal-eq} is shown (dotted line). The half-length of the tube is set to $L= 100~\mathrm{Mm}$, the thread half-length $l = 5~\mathrm{Mm}$, the coronal sound speed $c_\mathrm{sc}= 200~\mathrm{km\,s^{-1}}$, and the contrast $\chi = 100$.}
\label{wvsr_m1-fig}
\end{figure}

Similarly, we obtain the dispersion relation for the antisymmetric modes by eliminating the amplitude constants $B_2$ and $D_1$, and considering $B_1=0$
\begin{eqnarray}
\Omega \, \mathrm{cotg}&&\left[\Omega \Phi (1-W) \right]= \frac{1}{W\Phi} +
\frac{\chi W \Phi}{3} \left(\gamma +1 -\Omega^2\right) \nonumber \\
&&\times \frac{M\left( \frac{3}{2}+\frac{1-\Omega^2}{2\gamma},
\frac{5}{2}; \frac{\chi \gamma W^2 \Phi^2}{2}\right)}{
 M\left( \frac{1}{2} + \frac{1-\Omega^2}{2\gamma},\frac{3}{2}; 
\frac{\chi \gamma W^2 \Phi^2}{2}\right)} ~,
\label{drm1ant-eq}
\end{eqnarray}
in terms of the dimensionless variables of Equation (\ref{dim_units}).

In Figure~\ref{wvsr_m1-fig}, we have plotted the fundamental mode and several
overtones of the dispersion relations (Eqs. \ref{drm1sym-eq} and
\ref{drm1ant-eq}) as a function of the dip radius of curvature $R$. The fundamental
mode (solid line) decreases with $R$, clearly indicating that the
fundamental mode is affected by the dip curvature. In contrast, the frequencies
of the overtones are independent of $R$, indicating that these overtones are
purely sound-like or pressure-driven modes. We have also plotted the curve $\omega = \omega_{g}$ (or equivalently $\Omega= 1$), which corresponds to the case where the restoring force of the thread oscillation is exclusively the gravity projected along the tube \citep{luna2012a}. We see that the fundamental mode is very similar to the gravity-driven oscillation $\omega\approx \omega_\mathrm{g}$ for small values of $R$. For an intermediate radius of curvature, $R=200~\mathrm{Mm}$, the gravity-driven approximation
differs by $18\%$ with respect to the exact solution, and by $38\%$ for a relatively large radius $R=600~\mathrm{Mm}$. Therefore the approximation $\omega\approx \omega_\mathrm{g}$ is good for small and intermediate values of $R$. The increasing deviation from the purely gravity-driven case indicates that the pressure force has some influence for intermediate values of $R$, and becomes important for large radii.

The argument in the CHFs of the symmetric dispersion relation (Eq.
\ref{drm1sym-eq}) is small ($\chi \gamma W^2 \Phi^2/2 \ll 1$) for the values of
the parameters considered here. Additionally, $\Omega\Phi = \omega
L/c_\mathrm{sc} < 1$ for the fundamental mode. Thus, we expand the dispersion
relation and obtain   
\begin{equation}\label{wapad-eq}
\Omega^{2}  = 1+\frac{1}{W (1-W) \, \chi \, \Phi^{2}}~.
\end{equation}
Substituting the dimensional variables (Eq. \ref{dim_units}) yields an approximate expression for the
fundamental mode
\begin{equation}\label{wap-eq}
\omega_\mathrm{fund}^{2}=\frac{g_{0}}{R}+ \frac{c_\mathrm{sc}^{2}}{l \left( L - l \right)\chi}~.
\end{equation}
The fundamental frequency has two contributions: the gravity-driven frequency $\omega_{g}$, and the pressure-driven slow oscillation frequency
\begin{equation}\label{w-pressuredriven-eq}
\omega_\mathrm{s}=\sqrt{\frac{c_\mathrm{sc}^{2}}{l \left( L - l \right)\chi}}~.
\end{equation}
Now Equation (\ref{wap-eq}) can be written as 
\begin{equation}\label{wtotal-eq}
\omega_\mathrm{fund}^{2}=\omega_{g}^{2}+\omega_{s}^{2}~.
\end{equation}

Figure (\ref{wvsr_m1-fig}) exhibits a perfect match between the approximate expression and the exact values of the fundamental mode frequencies for a range of $R$ values. The $\omega_\mathrm{g}$ and $\omega_\mathrm{s}$ terms are identical at $R\approx330~\mathrm{Mm}$, while for larger values of $R$ the major contribution to the restoring forces is the pressure gradient. However, the fundamental mode differs significantly from $\omega_\mathrm{s}$, except at huge values of $R$. \citet{oliver1993} and \citet{oliver1995} studied the oscillation modes of a
slab in a Kippenhahn-Schl\"uter magnetic configuration with a very slight curvature. They concluded that the forces responsible for the slow modes are the pressure gradients, and the frequency of the fundamental mode is well described by $\omega_{s}$, consistent with the present result.

Similarly to the symmetric case we find an approximate expression for the frequency of the first overtone. In the range of small $l$ and large $R$ the term $W \Phi$ is small in the antisymmetric dispersion relation (Eq. \ref{drm1ant-eq}). The only way to balance the first term of the right-hand side of the equation is for the argument of the cotangent function to be near the first asymptote; then
\begin{equation}\label{1stovertoneapprox-eq}
\Omega \approx \frac{\pi}{(1-W) \Phi}~.
\end{equation}
In dimensional variables this equation is simply
\begin{equation}
\omega=\frac{\pi c_\mathrm{sc}}{L-l}~,
\end{equation}
which is the frequency of a standing wave trapped in one of the hot sections of the tube with wavelength $2(L-l)$. In Figure \ref{wvsr_m1-fig} we see that the frequency is more or less independent of $R$.

\begin{figure}[h!]
  \center{\resizebox{\hsize}{!}{\includegraphics{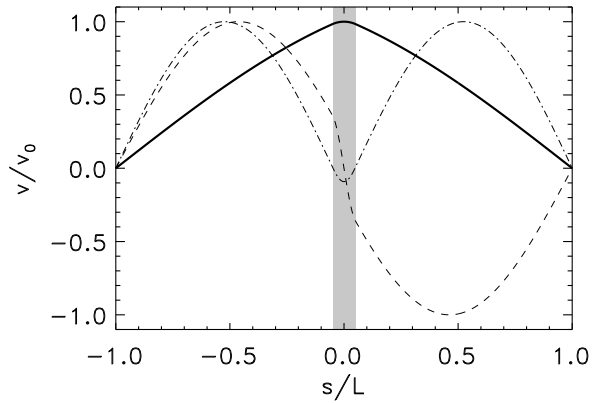} }}
\caption{The normalized velocity along the flux tube, $v\left(s\right)$, for the fundamental mode (solid line), the first antisymmetric overtone (dashed line), and the first symmetric overtone (dot-dashed lines). The flux tube half-length is $L = 100~~\mathrm{Mm}$, the thread half-length is $l = 5~\mathrm{Mm}$, dip radius of $R=75~\mathrm{Mm}$, coronal sound speed $c_\mathrm{sc}=200~\mathrm{km s^{-1}}$, and contrast $\chi=100$.}
\label{spatialdistmodes-fig}
\end{figure}

In Figure~\ref{spatialdistmodes-fig} the velocity perturbations of the first
three normal modes are shown. In the fundamental mode the maximum
velocity is centered within the thread and the
thread speed is quite uniform, indicating that motion resembles the motion of a solid
body. In the coronal part the velocity reduces to zero at the footpoints
as dictated by the boundary conditions. The first overtone is an
antisymmetric mode with a node ($v=0$) at the center of the tube. In this mode
the motion consists of compressions and rarefactions of the cool plasma with no net
displacement of the thread. The second overtone has a node
at each thread end, so the motion of the plasma is complex in this mode. All
overtones reach their maximum velocities in the coronal parts of the tube. 

\begin{figure}[ht]
  \center{\resizebox{\hsize}{!}{\includegraphics{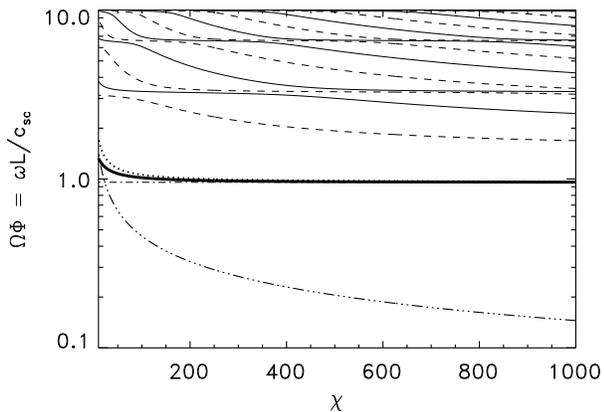} }}
\caption{Same as Figure \ref{wvsr_m1-fig} for the symmetric (solid lines) and antisymmetric (dashed lines) normal modes for Model 1 as a function of the density contrast $\chi$ at the interface between the cool and hot plasma. The fundamental mode is plotted with a thick solid line. The half-length of the tube is set to $L= 100~\mathrm{Mm}$, the thread half-length $l = 5~\mathrm{Mm}$, a coronal sound speed $c_\mathrm{sc}=200~\mathrm{km\,s^{-1}}$, and the radius of curvature $R=75~\mathrm{Mm}$.}
\label{wcontrast-fig}
\end{figure}

The frequencies of the symmetric and
antisymmetric normal modes as a function of the contrast $\chi$ are shown in Figure \ref{wcontrast-fig}. The fundamental mode rapidly reaches a constant value for increasing values
of $\chi$, reflecting a weak dependence on the contrast. The frequency of the
approximation $\omega_\mathrm{fund}$ is very similar to the fundamental mode.
For relatively large values of the contrast the match between both curves is
very good. The frequency $\omega_\mathrm{s}$ is inversely proportional to $\chi$ and thus rapidly goes to zero (triple dot-dash line in Fig. \ref{wcontrast-fig}), so the main contribution to the fundamental mode (Eq. \ref{wtotal-eq}) comes from the gravity term $\omega_g$. Then, for sufficiently large values of $\chi$, the oscillation of the thread decouples from the hot plasma and is governed by the gravity forces as shown in \S \ref{uncoupled-sec}. The frequencies of the overtones have also a complex dependence on $\chi$, with {\em avoided crossings} between modes. These avoided crossings are similar to those found in \citet{diaz2002} and \citet{diaz2006}, and have the same origin: two modes cannot have the same frequency, since it would violate the conditions in a St\"{u}rm-Liouville problem such as our Equation (\ref{waveadim-eq}). Furthermore, the modes can have extrema
in the prominence region and the hot region, and an avoided crossing occurs 
when the amplitude of the extrema in the prominence region changes from being
small (such as the first overtone in Fig. \ref{spatialdistmodes-fig}) to being
larger than the amplitude in the hot region.

\section{Influence of curvature in the hot regions}\label{curved-hot-sec}

We consider next the effects of curvature in the hot parts of the tube, 
which were assumed to be straight in the previous section. Two different cases are discussed.

\subsection{Model 2: concave-up field in the hot region}\label{model2-sec}

\begin{figure}[h!]
  \center{\resizebox{\hsize}{!}{\includegraphics{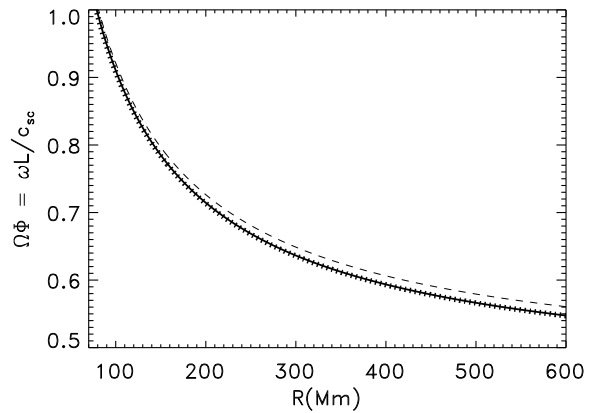} }}
\caption{Same as Figure \ref{wvsr_m1-fig} for the frequency of the fundamental mode as a function of $R$, for Model 1 (dotted line), Model 2 (solid line), and Model 3 (dashed line). In Model 3 we have condisered a dip half-length $d=20~\mathrm{Mm}$, and a radius of curvature of the legs $R_\mathrm{2}=-100~\mathrm{Mm}$.}
\label{model-comparison-fig}
\end{figure}

We consider a tube with constant curvature in the entire domain (see middle sketch of
Fig. \ref{sketch}). We assume that isothermal cool plasma occupies the region
$|s|\le l$, while the rest of the tube is filled with
isothermal hot plasma. In this model we have the limitation that $|R| \gg L$ in order
to satisfy the small-angle approximation $|s/R| \ll 1$ (see  \S \ref{tubes-curv-sec}). Thus
we can only consider the range of large curvature radii. As in the previous section (\S \ref{model1-sec}) we have found the dispersion relation for the normal modes of this model. Figure \ref{model-comparison-fig} shows that the fundamental frequency is essentially identical to that of Model 1. The frequencies and spatial distribution of the velocity obtained with both symmetric and antisymmetric dispersion relations are very similar to the corresponding solutions of Model 1 (\S \ref{model1-sec}) for all values of $R$, $L$, $\chi$, and $l$. For this reason we do not show the dispersion relations for Model 2.

\subsection{Model 3: M-shaped flux tube}

Finally, we consider a more realistic model composed of three different regions (see bottom sketch of Fig. \ref{sketch}). The dipped part is equivalent to Model 2,  with the cool and dense thread of length $2d$ centered and filling a portion  $2l$, and the remaining portion of the dip is filled with hot rarefied plasma. This dipped part is connected to the chromosphere by two concave-down segments filled with hot plasma. The resulting flux tube has an ``M'' shape, similar to the thread-bearing flux tubes identified by \citet{luna2012}.

The new feature of this M-shaped tube is the concave-down region, where the radius of
curvature is negative. Thus in this region the linear substitution in Equation (\ref{chn_var})
must be replaced by
\begin{equation}
r=s \left( \frac{-\gamma g_{0}}{2 c_\mathrm{s}^{2} R_2} \right)^{1/2},
\label{chn_var2}
\end{equation}
with $R_2 < 0$, the radius of curvature of the concave-down region. With this new definition Equation (\ref{wavegravity-eq}) becomes
\begin{equation}\label{waveadim-eq2}
 \frac{\partial^{2} v}{\partial r^{2}} +2 r \frac{\partial v}{\partial r}  -
 \frac{2}{\gamma}\left(\frac{R_2 \omega^2}{g_0} -1 \right) v =0.
\end{equation}
The solution of this equation can be written again in terms of CHFs as
\begin{eqnarray}
v (r) &=& A_1 e^{-r^2} \, M \left( \frac{2+\lambda}{4}, \frac{1}{2}; 
	r^2 \right)  \nonumber \\
&+& A_2 e^{-r^2} \, r \, M \left( \frac{4+\lambda}{4}, \frac{3}{2}; 
	r^2 \right)~.
\label{gen_sol2}
\end{eqnarray}
This functional form must be used for $d<|s|\le L$, while for $|s| \le d$, Equation (\ref{gen_sol}) is still valid. Using the boundary conditions to match the solutions at $s=l$, $s=d$ (Eq. \ref{jump_cond}) and the line-tying boundary condition at $s=L$ (Eq. \ref{line-tying}), we obtain a very cumbersome dispersion relation which involves products of five CHFs. For this reason, this dispersion relation is not shown in here. 

Figure \ref{model-comparison-fig} shows that the fundamental frequency for Model 3 is very similar to that of Model 1. In this figure we have considered a radius of curvature in the flux-tube legs of $|R_2|=100~\mathrm{Mm}$ (note that this is the smallest possible value allowed by the small angle condition, $|R_{2}| > L$). For larger values of $|R_2|$ the difference between the frequencies of Models 1 and 3 are even smaller, and zero when $|R_2|\to \infty$. We have also studied the dependence of the frequencies on the other parameters of the system and found identical results.

\section{The restoring force}\label{restoring-sec}

In our theoretical model of prominence oscillation there are two restoring forces: the gas pressure and the gravity. Their relative importance is determined by the ratio 
\begin{equation}\label{ratiosoundgravity-eq}
\frac{\omega_\mathrm{s}^{2}}{\omega_\mathrm{g}^{2}}=\frac{R c_{sc}^{2}}{l (L -l)
\chi g_{0}} =\frac{R}{R_\mathrm{lim}}
\end{equation} 
where $R_\mathrm{lim}= l (L- l) \chi g_{0} / c_{sc}^{2}$ is a reference radius of curvature determined by measurable filament properties. The restoring force of the LAL oscillation is mainly the solar gravity when this ratio is small, i.e., $R \ll R_\mathrm{lim}$. In contrast, the oscillation is pressure-driven for a large value of the ratio or $R \gg R_\mathrm{lim}$. With the data shown in Figure~\ref{wvsr_m1-fig} we find that $R_\mathrm{lim}= 325~\mathrm{Mm}$, so for smaller radii of curvature the gravity is the restoring force. The length of the threads and density contrast have a wide range of measured or estimated values \citep[see, e.g.,][]{labrosse2010,mackay2010}. Considering larger values of the thread length and contrast of $l = 10~\mathrm{Mm}$ and $\chi = 200$, we find $R_\mathrm{lim}= 1200~\mathrm{Mm}$, indicating that the gravity dominates even for almost flat tubes. \citet{zhang2012} studied the observed longitudinal oscillations of a prominence, and found a small radius of curvature less than $100~\mathrm{Mm}$.
Therefore the restoring force is certainly the gravity, as we would expect in those structures. Additionally, the relative importance of the two restoring forces also depends on the geometrical parameter $R/L$ (Eq. \ref{ratiosoundgravity-eq}), which can be determined from the global geometry of the filament magnetic structure. For the deep-dip flux tubes of a double sheared arcade \citep{devore2005}, this factor is always $R/L<1$ \citep{luna2012}.

In \S \ref{uncoupled-sec} we discussed that the spatial coherence of the motion of the thread is determined by $\lambda$. Combining Equation (\ref{lambda_def}) with Equation (\ref{wtotal-eq}), we find that in Models 1 to 3
\begin{equation}\label{lambda_sound-eq}
\lambda=\frac{2}{\gamma}\frac{\omega_\mathrm{s}^{2}}{\omega_\mathrm{g}^{2}}~.
\end{equation}
Thus the relative importance of the pressure and gravity forces determines the way the thread moves. For a gravity-driven oscillation $\lambda=0$, the thread basically moves almost as a solid body. However, for a pressure-driven motion the thread compresses and rarefies, changing its shape in the oscillation. For the reference values of Figure~\ref{wvsr_m1-fig} and a radius of $75~\mathrm{Mm}$ the coherence parameter is $\lambda=0.3$. The velocity profile for this parameter value plotted in Figure~\ref{chf-plots-fig}, is quite flat in the thread, indicating that the motion is very similar to the displacement of a rigid solid body. 

\section{Discussion and conclusions}

In this work we have studied the influence of the curvature on the longitudinal oscillations of a prominence. We have considered three different models, in which the region where the thread resides is modeled as a tube segment with uniform curvature $R$. The differences between the three models are in the hot regions of the tubes. We have found that the frequency of the fundamental mode is dependent on $R$ and can be approximated by $\omega_\mathrm{fund}^2 = \omega_\mathrm{g}^2 + \omega_\mathrm{s}^{2}$, where $\omega_\mathrm{g}$ and $\omega_\mathrm{s}$ are the frequencies of the gravity-driven pendulum and the fundamental slow mode of a straight tube respectively. For small and intermediate $R$ the frequency is very close to $\omega_\mathrm{g}$. We modeled a prominence with realistic dimensions and found most of the filament flux-tubes have dips with a radius of curvature around $75~\mathrm{Mm}$ \citep{luna2012,luna2012a}. We also inferred the radius of curvature from the oscillation periods reported by \citet{jing2003} and \citet{vrsnak2007}, and found $152~\mathrm{Mm}$ and $62~\mathrm{Mm}$ respectively. \citet{zhang2012} observationally determined the radius of filament-dip curvatures to be less than $100~\mathrm{Mm}$. Thus, the observations and theoretical models are consistent with this range of small and intermediate $R$. For larger radii the pressure force becomes more important and the fundamental frequency differs from $\omega_\mathrm{g}$. The frequencies of the overtones are basically independent of the curvature of the tube, consistent with the slow nature of these modes. 

The fundamental mode is also weakly dependent on the density contrast, $\chi$, for small contrast, but as $\chi$ increases the fundamental mode rapidly reaches a constant value that coincides with $\omega_\mathrm{fund}$ or $\omega_\mathrm{g}$ (note that $\omega_\mathrm{fund} \approx \omega_\mathrm{g}$ because $\omega_\mathrm{s}$ is very small for relatively large contrast). Therefore the oscillation of the thread decouples from the environment for relatively larger values of the contrast, and the thread oscillates with the frequency $\omega_\mathrm{g}$. Observational estimates of prominence densities give a broad range of possible $\chi$ values, but typically the density contrast is 100 or larger \citep[see review by][]{labrosse2010}. Therefore the threads are in the range of large density contrasts. The overtones depend on $\chi$, indicating the sound-like nature of these modes. The spatial velocity distribution of the fundamental mode along the tube is symmetric with respect to the tube center; although the maximum is located at the thread center, the velocity in the thread is more or less uniform. In the fundamental mode the motion is mainly concentrated in the thread. The overtones produce complex compression and rarefaction motions with small or zero net displacements of the thread.

In Models 2 and 3, with curved hot regions, the frequencies and the spatial distribution are very similar to the corresponding values for Model 1. We conclude, therefore, that the shape of the hot regions is irrelevant to the longitudinal oscillation, and that using a straight field-line approximation in these regions gives results that are accurate enough and much easier to compute. Hence, the approximation given in Equation~(\ref{wtotal-eq}) is quite robust, despite only being truly valid for Model 1. The sound speed in the corona is very high, and the contribution of the first term in Equation (\ref{pde_curv}) is larger than terms involving curvature. Thus, the resulting motion in the coronal parts of the flux tube is well described by Equation (\ref{straightwave_eq}).

The relative importance of the pressure and gravity forces is determined by the geometry of the dipped part of the filament flux tubes. We found that the oscillation is gravity driven when $R\ll R_\mathrm{lim}$, where $R_\mathrm{lim}$ is determined by the properties of the filament flux tubes and the cool thread. For typical prominences $R_\mathrm{lim}$ is much larger than the radius of curvature of filament models or observationally inferred values.

We conclude that the longitudinal oscillations of a prominence are strongly influenced by the curvature of the dipped magnetic fields. In the fundamental mode the gravity dominates for small and intermediate radius of curvature, consistent with values in our multi-threaded prominence model \citep{luna2012} and with the values derived from observed oscillations. Similarly, for relatively large density contrast between the prominence and the corona the main restoring force is also the gravity. Thus, the frequency of the LAL oscillations is given by $\omega_\mathrm{g}=\sqrt{g_{0}/R}$, and the pressure forces introduce only a small correction, showing that the LAL oscillations are not pure slow modes. This demonstrates that the $\omega_\mathrm{g}$ expression is robust and can be used for seismology of prominences, as we showed in \citet{luna2012a}.

In this work we have not studied the damping of the observed LAL oscillations. In \citet{luna2012a} we found that the damping is associated with mass accretion onto the threads and nonadiabatic effects. The former produces strong damping at the beginning of the oscillation, and the latter yields a weak damping throughout the oscillation. In order to have a full and self-consistent model of prominence oscillations we must perform a nonlinear study, including the temporal variation of the prominence mass and the nonadiabatic effects. This will be the subject of a future work.

\acknowledgements

This work has been supported by the NASA Heliophysics SR$\&$T program. M.L. also acknowledges support from the University of Maryland at College Park and the people of CRESST. A.J.D. acknowledges the financial support by the Spanish Ministry of Science through project AYA2010-18029.

\end{document}